\documentclass[10pt,conference]{IEEEtran}
\pdfoutput=1
\usepackage{diagbox}
\usepackage{graphicx}
\usepackage{amsmath}
\usepackage{amsfonts}
\usepackage{algpseudocode}
\usepackage{algorithm}
\usepackage{subfigure}
\usepackage{tikz}
\usepackage{epsfig}
\usepackage{cite}
\usetikzlibrary{shapes}
\usetikzlibrary{arrows,automata}
\usetikzlibrary{positioning}
\usetikzlibrary{patterns}
\usetikzlibrary{backgrounds}
\newtheorem{theorem}{Theorem}

\newtheorem{definition}{Definition}
\newtheorem{lemma}{Lemma}

\newtheorem{example}{Example}

\usepackage[firstpage]{draftwatermark}
\SetWatermarkText{\hspace*{7.2in}\raisebox{9.2in}
  {\includegraphics[scale=0.1]{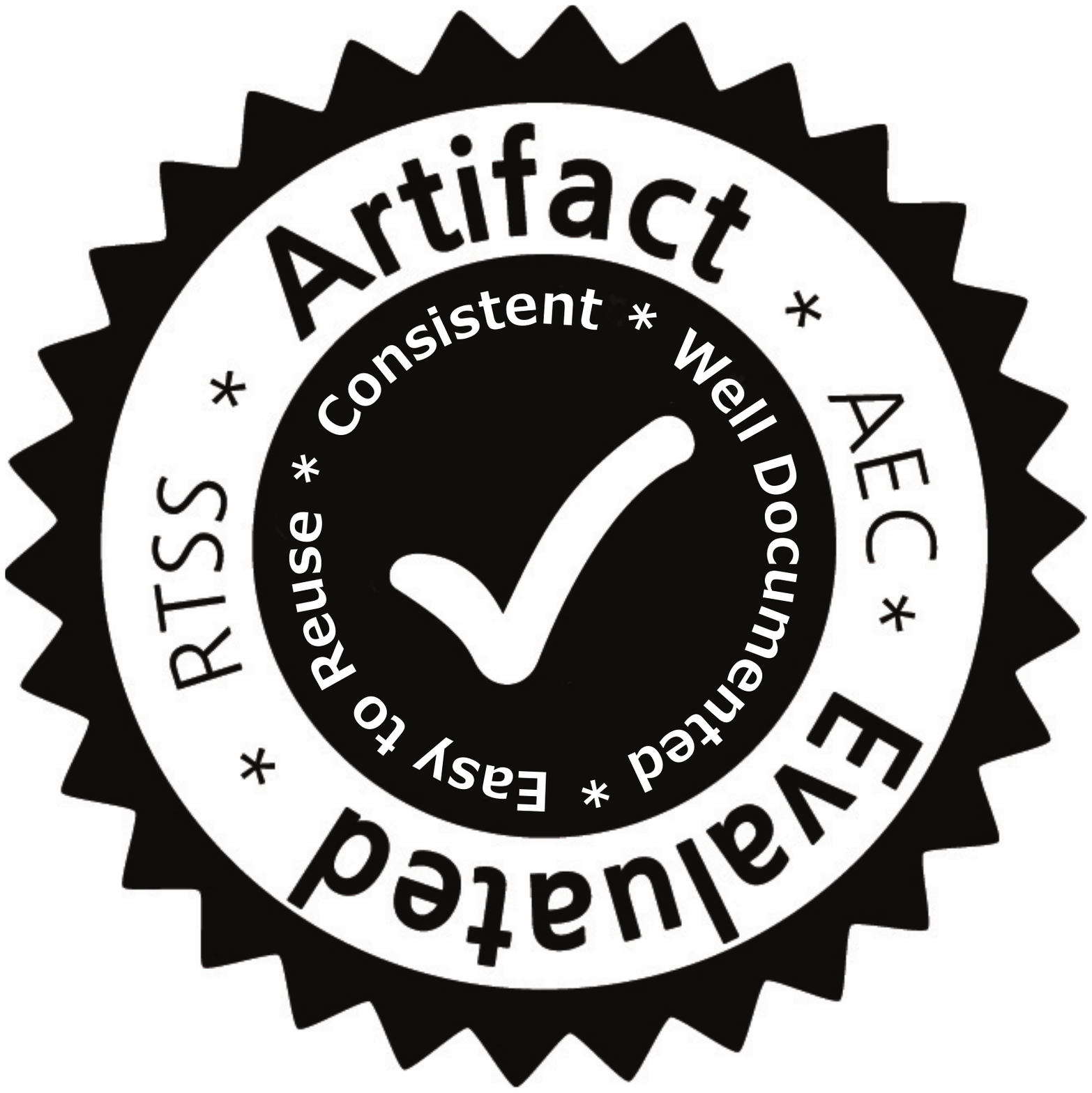}}}
\SetWatermarkAngle{0}

\begin{document}

\title{Dynamic Budget Management with Service Guarantees for Mixed-Criticality Systems\thanks{This paper has passed an Artifact Evaluation process.}}
\author{
\IEEEauthorblockN{Xiaozhe Gu, Arvind Easwaran}
\IEEEauthorblockA{Nanyang Technological University, Singapore \\
Email: guxi0002@e.ntu.edu.sg, arvinde@ntu.edu.sg}
}
\maketitle

\begin{abstract}
Many existing studies on mixed-criticality (MC) scheduling assume that low-criticality budgets for high-criticality applications are known apriori. These budgets are primarily used as guidance to determine when the scheduler should switch the system mode from low to high. Based on this key observation, in this paper we propose a dynamic MC scheduling model under which low-criticality budgets for individual high-criticality applications are determined at runtime as opposed to being fixed offline. To ensure sufficient budget for high-criticality applications at all times, we use offline schedulability analysis to determine a system-wide total low-criticality budget allocation for all the high-criticality applications combined. This total budget is used as guidance in our model to determine the need for a mode-switch. The runtime strategy then distributes this total budget among the various applications depending on their execution requirement and with the objective of postponing mode-switch as much as possible. We show that this runtime strategy is able to postpone mode-switches for a longer time than any strategy that uses a fixed low-criticality budget allocation for each application. Finally, since we are able to control the total budget allocation for high-criticality applications before mode-switch, we also propose techniques to determine these budgets considering system-wide objectives such as schedulability and service guarantee for low-criticality applications.
\end{abstract}
\section{Introduction}

An increasing trend in safety-critical real-time applications is that multiple functionalities with different levels of ``criticality" (importance) are integrated together on a single computing platform~\cite{prisaznuk1992integrated}. To efficiently share the computing platform among those applications while ensuring isolation between different criticalities, Vestal proposed the classic mixed-criticality (MC) task and scheduling model~\cite{Ves07}. This task model is an extension of the standard sporadic real-time task system for two criticality levels. A task is defined as either a high-critical (HC) or a low-critical (LC) task.  A HC task $\tau_i$ has two execution time estimates $C_i^L$ and $C_i$. While $C_i$ is assumed to be greater than or equal to the worst case execution time (WCET) of the task, $C_i^L$ is a lower estimate ($C_i^L \leq C_i$) that may not be sufficient for some jobs of the task. A LC task $\tau_i$ only has a single execution time estimate $C_i$ that is assumed to be greater than or equal to its WCET. 

Under the classic MC scheduling model, schedulability of HC tasks is assessed under the standard assumption that no task would execute beyond $C_i$. On the other hand, schedulability of LC tasks is only assessed under the assumption that each HC task $\tau_i$ would not execute beyond $C_i^L$. Consequently, a fundamental difference between this model and the standard non-MC scheduling model is that the scheduler can prioritize HC tasks over LC tasks when additional processing capacity ($> C_i^L$) is required for them. The system can then be seen as being in two different execution modes at runtime; LC mode as long as no job of any HC task $\tau_i$ executes beyond $C_i^L$, and HC mode thereafter during which no LC deadlines are required to be met.

From the above discussion we can observe that the execution estimate $C_i^L$ is primarily used as a ``budget'' for the HC task $\tau_i$ in the LC mode. In fact, it helps the scheduler to determine whether the system should switch to the HC mode. Let $B_i^L$ denote the maximum budget allocated to jobs of $\tau_i$ by an MC scheduler in the LC mode. Then, we can see that the classic MC scheduling model uses a static (fixed) budget allocation in the LC mode; $B_i^L = C_i^L$ for HC tasks and $B_i^L = C_i$ for LC tasks. Hence, we denote this model as the \textbf{static model}. For further details like execution semantics and definition of schedulability for the static model, please refer to previous works e.g.,  \cite{Ves07,BBD11,BBA12}.

\textbf{Motivation.} In this paper we propose a more dynamic MC task and scheduling model  (denoted in short as \textbf{dynamic model}) on \textbf{a uniprocessor platform} based on the following principle. HC tasks can be allocated budgets in the LC mode dynamically at runtime depending on the overall processing requirements of the task system. As long as the allocated budgets continue to ensure schedulability requirements for all the tasks, it should be safe to do so.

One important advantage of the dynamic model over the static one is that the application designers are not required to specify $C_i^L$ any more, thus reducing their burden. But then a problem arises as to how a scheduler can safely determine mode-switch in the dynamic model so that schedulability of HC tasks is guaranteed at all times. To address this problem we propose a technique that combines offline schedulability analysis with runtime budget allocation strategy for HC tasks. Offline, we use schedulability analysis to determine a \textbf{total LC budget} allocation (single value) for all HC tasks combined. At runtime, we use a strategy to allocate budgets to individual jobs of HC tasks based on this total budget. Thus, the budget computed offline is used as an indicator to determine mode-switch in the dynamic model, and hence the schedulability of HC tasks is ensured.

A system-wide budget for HC tasks in the LC mode, as opposed to budgets for individual tasks as in the static model, allows more flexibility at runtime to distribute this budget depending on execution requirements of individual jobs. It thus eliminates one of the pessimistic assumptions made in many existing MC studies, which is that when a single HC job has a high execution requirement (e.g., beyond $C_i^L$), all the other HC jobs in the system would also have a high execution requirement in the near future.  

As a result of this dynamic budget allocation for individual HC tasks, the mode-switch from LC to HC can also be potentially postponed in the new model. Delaying this mode-switch has a significant implication on the service provided to LC tasks, because many existing MC scheduling strategies either completely drop LC tasks or offer degraded service to them after the mode-switch.


Note that in the proposed dynamic model, even though HC tasks have a combined budget allocation in the LC mode, each HC task $\tau_i$ must still be provided as much budget as it needs upto its WCET at all times (i.e., $C_i$ units).    Hence this new model has the same fault-isolation properties as systems in use today (e.g. mixed-criticality systems in avionics and automotive), and therefore we believe the model has a strong practical relevance.


\textbf{Contributions.} The contributions of this paper can be summarized as follows.

\textbf{1)~Dynamic MC task and scheduling model (Section~\ref{sec:new})}: We propose a new MC task and scheduling model in which jobs of HC tasks are allocated budgets in the LC mode dynamically at runtime depending on their execution requirements. To ensure schedulability for HC tasks, we use a system-wide budget allocation that is determined offline as a guidance for mode-switch.

\textbf{2)~Runtime budget allocation strategy (Section~\ref{sec:dynamic})}: We propose a runtime technique to distribute budgets to jobs of HC tasks in the LC mode, depending on the execution requirement of jobs as well as the total budget allocation determined offline. We also prove that under certain conditions the proposed runtime strategy is optimal in terms of being able to postpone the mode-switch as much as possible.

\textbf{3)~Determination of total budget for HC tasks (Section~\ref{sec:test})}: We propose an offline technique to determine the total LC budget allocation for HC tasks. This technique is based on uniprocessor schedulability analysis for a variant of the well known EDF-VD (Earliest Deadline First with Virtual Deadlines) MC scheduling policy~\cite{BBA12}. 

\textbf{4)~Minimum service guarantee for LC tasks in the HC mode (Section~\ref{sec:test})}: 


We  also propose a strategy that enables LC tasks to receive budgets after the system switches mode. Similar to the total LC budget for HC tasks, these budgets for LC tasks are also determined offline based on schedulability analysis, and hence guarantee minimum service to LC tasks in the HC mode. Since we compute budgets for both HC and LC tasks in the dynamic model, we are able to trade-off the service guarantee for LC tasks in the HC mode against the budget reserved for HC tasks in the LC mode. 

\textbf{5)~Scheduling policy EDF-UVD (Section~\ref{sec:schedule})}: We propose a MC scheduling policy called Earliest Deadline First - Universal Virtual Deadlines (EDF-UVD), which is similar to EDF-VD, except that in addition to HC tasks, even LC tasks have virtual deadlines under EDF-UVD. This policy is useful in systems with guaranteed service for LC tasks. By using virtual deadlines for LC tasks, we are able to differentiate between the amount of execution that is guaranteed to LC tasks in all modes, versus the additional amount that is guaranteed only in the LC mode.    

\section{Related Work}
\label{sec:related}

The static model~\cite{Ves07} is widely used in many previous studies for representing MC real-time workloads. This model requires multiple execution time estimates for different criticality levels in a system. Therefore a task may have up to five execution time estimates in a five-level MC system. As pointed out by Burns~\cite{AB15}, it is an undue burden for the application designer to obtain such multiple execution time estimates. In~\cite{AB15}, Burns also proposes a simplification to the static model, where each task only has two execution time estimates. By contrast, in this paper we propose using a single estimate from the application designer.

Many existing work on MC scheduling (e.g.,~\cite{BBD11,BaFo11,GES11,BBA12,EkYi12,Eas13}) share the pessimistic strategy that all LC tasks will be immediately dropped once the system switches to HC mode. Other studies have presented solutions to improve support for LC tasks~\cite{BuBa13,JZP13,HND14,Su13,GU15,PCH13,PCH14,fleming2014incorporating,ren2015mixed,Weaklyhard}. These solutions can be broadly categorized into two classes.

\textbf{1)} The first category of studies~\cite{BuBa13,JZP13,PCH14,HND14,Su13,Weaklyhard,SAZ16} support LC tasks by offering a degraded (and in some cases guaranteed) service to all of them when the system is in the HC mode. They do this either by reducing the dispatch frequency of jobs or by executing the LC tasks as background (low-priority) workload.     

\textbf{2)} The second category of studies~\cite{PCH13,GU15,ren2015mixed,fleming2014incorporating} support LC tasks by offering a degraded service to only a subset of them, while keeping the service to others intact, depending on the specific HC tasks that demand additional execution. However, these studies do not provide any minimum guaranteed service to the LC tasks in the HC mode. 

None of the above studies considered a dynamic budget allocation model for HC tasks such as the one proposed in this paper. Further, the LC service strategy we propose belongs to both the categories described above. It offers a minimum guaranteed service to all LC tasks in the HC mode (similar to some of the studies in the first category), and at the same time we are able to control the mode-switch depending on individual job execution requirements of HC tasks (similar to the second category). Thus, by combining both these approaches we are able to explore the trade-off between LC service guarantee on one hand and mode-switch on the other.

There are few studies that focus on LC executions in the HC mode. Bailout protocol~~\cite{BateBD15} reduces the negative impact on LC tasks via a timely return to LC mode. Another study~~\cite{guo2015edf}  uses the probability that $C_i^L$ would be exceeded in the static model and derives corresponding schedulability analysis with permitted system failure probability. These studies are orthogonal to the problems addressed in this paper.

To protect HC tasks from overload conditions in LC tasks, the work in \cite{neukirchner2013multi} proposes runtime techniques to monitor and safely switch between feasible activation patterns of LC tasks. In contrast, we focus on runtime allocation of execution time budgets to HC tasks with the objective of improving support for LC tasks. Although there is some similarity between the two runtime mechanisms in that they both allocate slack to tasks, the task models, objectives and strategies are all different. Further, as opposed to activation bounds for individual tasks in \cite{neukirchner2013multi}, we are able to use properties of the runtime strategy in schedulability analysis to derive a single budget for all HC tasks combined. As a result, we have more flexibility to allocate this budget at runtime, and are also able to show that our runtime strategy is optimal among all fixed-budget strategies in terms of the ability to delay mode-switch.
\section{Dynamic MC Task and Scheduling Model}
\label{sec:new}

\subsection{Task model}

In the dynamic MC task model, each task $\tau_i$ is defined as a tuple $(T_i,C_i,L_i\in\{LC,HC\})$, where $T_i$ is the minimum separation time between successive job releases, $L_i$ denotes the criticality level, and $C_i$ upper bounds the worst case execution time of the task. Thus the application designer only needs to provide one execution time estimate $C_i$ as in the standard non-MC task model. For a HC task $\tau_i$, its HC budget is fixed at $B_i^H=C_i$ similar to the static model, but its LC budget is determined at runtime. It varies depending on the past execution demand of all HC jobs (details in Section~\ref{sec:dynamic}). On the other hand, for a LC task $\tau_i$ its LC budget is fixed at $B_i^L=C_i$ similar to the static model, and its HC budget is also fixed at $B_i^H=C_i\times \alpha_i~(\alpha_i \in [0,1])$ where $\alpha_i$ is determined offline. $\alpha_i$ denotes the minimum service that task $\tau_i$ is guaranteed in the HC mode. We focus on implicit deadline task systems in this paper (i.e., relative deadline $D_i$ is equal to $T_i$ for each task), and consider the problem of scheduling $n$ such tasks $\tau = \{ \tau_1, \ldots \tau_n \}$ on a uniprocessor platform.

Let $u_i=\frac{C_i}{T_i}$ denote the utilization of task $\tau_i$, $\tau_L \in \tau$ denote the set of LC tasks and $\tau_H \in \tau$ denote the set of HC tasks. Also, let $U_H=\sum\limits_{\tau_i\in \tau_H}u_i$ and $U_L=\sum\limits_{\tau_i\in \tau_L}u_i$.

In each system mode (LC or HC), the \emph{service level} of a task depends on the amount of budget statically reserved for the task in the dynamic model. We define this service level based on the proportion of total execution requirement $C_i$ that is statically reserved in each mode.        
 
\begin{definition}[Task Service Level]
\label{def:def1}
For each task $\tau_i \in \tau_L$, its \emph{LC service level} is $\frac{B_i^L}{C_i} = 1$ and its \emph{HC service level} is $\frac{B_i^H}{C_i} = \alpha_i$. For each task $\tau_i \in \tau_H$, its \emph{HC service level} is $\frac{B_i^H}{C_i} = 1$. 
\end{definition}

{Note that, even though a total LC budget is statically reserved for  HC tasks  combined in LC mode, for an individual HC task, its LC budget  is not \textbf{statically} reserved.  Therefore we do not define LC service level for an individual HC task.} {Intuitively speaking, the LC/HC task service level characterizes the  proportion of utilization that is guaranteed to the task in LC and HC mode, respectively.}  Analogously, we can define the system service levels as follows. 


\begin{definition}[System Service Level]
\label{def:def2}
The \emph{HC system service level} of task set $\tau$ is denoted as $\alpha^*$ and defined as follows. 
\begin{equation}
\alpha^*=\frac{\sum_{\tau_i\in \tau_L}\alpha_i\times u_i}{U_L}
\end{equation}
{Here $\alpha^*$ characterizes the  proportion of $U_L$ that is statically reserved for all the LC tasks in HC mode.}  The \emph{LC system service level} of task set $\tau$ is denoted as $\beta^*$, and it is equal to the proportion of $U_H$ that is  reserved as total budget for all HC tasks combined in the LC mode. 
\end{definition}

As discussed in the introduction, $\beta^*$ is determined offline using schedulability analysis and considering several criteria such as HC system service level $\alpha^*$, average system service level in LC and HC modes, etc (see Section~\ref{sec:opt}). Note that the LC system service level only depends on HC tasks, whereas the HC system service level only depends on LC tasks. This is consistent with the definitions of task service levels, because LC tasks (likewise HC tasks) receive full service in the LC (likewise HC) mode.

Similar to the static model, we can also define a schedulability criteria for algorithms that schedule task systems based on the dynamic model presented above. Since LC tasks also receive guaranteed service in the HC mode ($\alpha^*$), unlike the static model, their deadlines cannot be ignored. Further, since individual HC tasks do not have a fixed budget in the LC mode, the scheduler only needs to ensure that each job of HC task $\tau_i$ receives as much execution as it needs up to $C_i$ in both modes. If it is able to provide a budget of $C_i$ for some job of $\tau_i$ while remaining in the LC mode, then it is free to do so.   In fact, this flexibility in budget allocation is a key advantage of the dynamic model.
\begin{definition}[MC-Schedulable]
A task system $\tau$ in the dynamic model is defined to be \textbf{MC-Schedulable} by a scheduling algorithm if the following two conditions hold:
\begin{enumerate}
\item Jobs of LC task $\tau_i$ with deadline in the LC mode receive up to $B_i^L (=C_i)$ units of budget each, and jobs of LC task $\tau_i$ with deadline in the HC mode receive up to $B_i^H (=\alpha_i \times C_i)$ units of budget each.
\item Jobs of HC task $\tau_i$ receive as much budget as they need as long as it does not exceed $C_i$.
\end{enumerate}
\end{definition}

\subsection{Runtime strategy for allocating LC budget to HC tasks}
\label{sec:dynamic}

In this section we present an efficient runtime strategy named MEBA (short for Maximum Execution-based Budget Allocation) for distributing the LC budget among jobs of HC tasks. While doing so we ensure that the total LC budget allocated to all HC tasks over any period of time is proportional to $\beta^* \times U_H$. This strategy also determines when a mode-switch will occur in the dynamic model; it is precisely when the total demand of all HC tasks exceed $\beta^* \times U_H$. To be able to do this efficiently, we need to store some information about budgets consumed by HC jobs in the past. We first use a simple example to illustrate MEBA, and then present the general approach. 

\begin{example}
Suppose a system $\tau$ has two HC tasks $\{\tau_1,\tau_2\}$. Consider the following sequence of job executions $\left< J_1^1, J_2^1,J_1^2 \right>$, where $J_i^j$ denotes the {$j_{th}$} job of task $\tau_i$. Before $J_1^1$ begins execution, we allocate a budget to it as follows. 
\vspace{-3mm}
\begin{small}
\begin{align*}
&\frac{b_1}{T_1} = \! \beta^*\times U_H \Rightarrow b_1 = T_1\times \beta^*\!\times U_H
\end{align*}	
\end{small}
Thus $J_1^1$ is allocated budget proportional to $\beta^* \times U_H$. If $J_1^1$ executes for $b_1$ time units but does not complete, the system will immediately switch to the HC mode. Otherwise after $J_1^1$ completes, the amount of execution that $J_1^1$ consumed (say $e_1^1 (\leq b_1)$) will be used to determine $b_2$, the budget for $J_2^1$. Thus before $J_2^1$ begins execution, we allocate a budget to it as follows. 

\vspace{-5mm}
\begin{small}
\begin{align*}
&\frac{e_1^1}{T_1}\!+\!\frac{b_2}{T_2} = \! \beta^*\!\times \!U_H 
\!\!\Rightarrow \!b_2 = \!T_2\times\! \left(\beta^*\times U_H\!-\!\frac{e_1^1}{T_1}\right)\\
\end{align*}
\end{small}
\vspace{-8mm}

A fraction of budget equal to $e_1^1/T_1$ is reserved for future jobs of task $\tau_1$. The remaining budget of $(\beta^*\times U_H-e_1^1/T_1)$ is allocated to $J_2^1$. Suppose $J_2^1$ gets preempted by $J_1^2$ after executing for $e_2^1 (\leq b_2)$ time units. Then, a new value for $b_1$, which is now the budget for job $J_1^2$, will be computed as follows.  
\begin{small}
\begin{align*}
&\frac{b_1}{T_1}\!+\!\frac{e_2^1}{T_2} = \! \beta^*\!\times \!U_H 
\!\!\Rightarrow \!b_1 = \!T_1\times\! \left(\beta^*\times U_H\!-\!\frac{e_2^1}{T_2}\right)\\
\end{align*}	
\end{small}
When $J_2^1$ resumes execution at a later time instant, then $b_2$ will again be updated depending on the amount of execution $e_1^2$ consumed by $J_1^2$. If $J_1^2$ executes for no more than $e_1^1$ time units, then $b_2$ does not change. Otherwise, it will decrease to ensure that the total allocation is proportional to $\beta^* \times U_H$.
\begin{align*}
\!b_2 = \!T_2\times\! \left(\beta^*\times U_H\!-\!\frac{\max \{e_1^1, e_1^2 \}}{T_1}\right)\\
\end{align*}
\vspace{-10mm}
\end{example}

Thus, at all times we need to store for each HC task $\tau_i$ the largest execution time of any job of that task in the busy interval. We reserve a budget proportional to this largest execution time for all future jobs of that task. The MEBA budget allocation strategy can be described as follows. 

\textbf{MEBA Runtime Strategy.} For each HC task $\tau_i$, initialize $e_i^m=b_i=0$, {where $e_i^m$ denotes the maximum amount of time for which any job of $\tau_i$ has executed in the \textbf{latest busy interval},  and $b_i$ denotes the budget allocated to the current job of $\tau_i$}. In each mode, the following steps will be executed in sequence. 
\begin{itemize}
\item In the LC mode:
\begin{enumerate}
 \itemsep=-1pt
\item If a job of some HC task $\tau_i$ will be allocated to the processor at the current time instant, then update $b_i=T_i\times \left(\beta^*\times U_H-\sum_{\tau_j \in \tau_H \setminus \tau_i}\frac{e_j^m}{T_j}\right)$ before executing this job .
\item If a job of some HC task $\tau_i$ executes for $b_i$ time units in total but does not complete, then trigger a mode-switch to the HC mode. Skip remaining steps.
\item Suppose a job of some HC task $\tau_i$ gets preempted or completes at the current time instant. Let $e_i$ denote the total execution time consumed by this job so far. Then, update $e_i^m=\max\left\{e_i^m,e_i\right\}$.


\end{enumerate}
\item In the HC mode:
\begin{enumerate}
\item After an \textbf{idle instant}, reset the mode to LC and set $b_i = e_i^m = 0$ for each HC task $\tau_i$.
\end{enumerate}
\end{itemize}

The following lemma records an important property of MEBA.

\begin{lemma}[Mode-switch condition]
\label{lemma:mode-switch}
Suppose a mode-switch to HC mode is triggered at some time instant $t^*$ in a \textbf{busy interval} under MEBA. For each HC task $\tau_i$ and each time instant $t \leq t^*$, let $e_{i,t}^m$ denote the maximum amount of time for which any job of $\tau_i$ has executed before $t$ in that busy interval. Then, for all $t \leq t^*$,
\begin{equation}
\label{eq:lc_condition} 
	\sum_{\tau_i\in \tau_H} \frac{e_{i,t}^m}{T_i} \leq \beta^*\times U_H.
	\vspace{-2mm}
\end{equation}
Further, $t^*$ is the earliest time instant in that busy interval such that there is a job of some HC task $\tau_k$ with execution requirement $e_k > e_{k,t^*}^m$, and 
\begin{equation}
\label{eq:modeconditon} 
	\sum_{\tau_i\in \tau_H} \frac{e_{i,t^*}^m}{T_i} = \beta^*\times U_H.
\end{equation}
\label{lem:ms}
\end{lemma}
\begin{IEEEproof}
Since a busy interval only has one mode-switch to HC mode under MEBA, $t^*$ is the only such mode-switch instant in the busy interval under consideration.  

Suppose Equation~\eqref{eq:lc_condition} does not hold, and $t$ denotes the earliest time instant where it fails. Let $J_k$ denote a job of HC task $\tau_k$ that was continuously scheduled in the interval $[t', t)$ for some $t' < t$. The value of $b_k$ was updated at $t'$ to $T_k\times (\beta^*\times U_H-\sum_{\tau_i \in \tau_H \setminus \tau_k} e_{i,t'}^m/T_i )$ and remains the same up to $t$. Further, since $t$ is the earliest time instant at which Equation~\eqref{eq:lc_condition} fails, we know that $e_{k,t'}^m \leq T_k\times (\beta^*\times U_H-\sum_{\tau_i \in \tau_H \setminus \tau_k} e_{i,t'}^m/T_i ) = b_k$. Further, $\forall \tau_i \in \tau_H \setminus \tau_k, e_{i,t}^m = e_{i,t'}^m$, because none of these tasks executed in the interval $[t',t)$. Then, since Equation~\eqref{eq:lc_condition} fails at $t$, we get,
\begin{small}
\begin{align*}
& e_{k,t}^m > T_k \left (\beta^*\times U_H-\sum_{\tau_i \in \tau_H \setminus \tau_k} e_{i,t}^m/T_i \right ) \iff\\
 & e_{k,t}^m > T_k \left (\beta^*\times U_H-\sum_{\tau_i \in \tau_H \setminus \tau_k} e_{i,t'}^m/T_i \right ) 
\iff e_{k,t}^m > b_k
\end{align*}	
\end{small}

Since $e_{k,t'}^m \leq b_k < e_{k,t}^m$ and $J_k$ is the only job executing in $[t',t)$, it must be the case that $J_k$ itself has executed for a total time of $e_{k,t}^m$ by $t$. This, combined with the fact that $J_k$ did not execute for more than $e_{k,t'}^m$ before $t'$, indicates that there is some time instant in $[t',t)$ when $J_k$ has completed $b_k$ units of execution but remains incomplete. By definition of MEBA, this would have triggered a mode-switch to HC mode at that time instant. This is impossible however, because $t^* (\geq t)$ is the only mode-switch in the busy interval. Thus we have shown that Equation~\eqref{eq:lc_condition} holds for all $t \leq t^*$.

Since a mode-switch is triggered at $t^*$, from the definition of MEBA we know that there is a job $J_k^*$ of some HC task $\tau_k$ that remains incomplete at $t^*$. By $t^*$, $J_k^*$ has executed for exactly $\lambda_{k,t^*}=b_k=T_k\times (\beta^*\times U_H-\sum_{\tau_i \in \tau_H \setminus \tau_k}e_{k,t^*}^m/T_j)$ time units. Therefore at $t^*$, we have  $\sum_{\tau_i\in \tau_H\setminus \tau_k} e_{i,t^*}^m/T_i+\lambda_{k,t^*}/T_k= \beta^*\times U_H$, and the total execution demand $e_k$ of $J_k^*$ is greater than $\lambda_{k,t^*}$. Then $e_{k,t^*}^m=\lambda_k^*$, because Equation~\eqref{eq:lc_condition} holds at $t^*$ and $e_{k,t^*}^m \geq \lambda_{k,t^*}$ by definition. Thus we have identified the conditions that must hold when a mode-switch is triggered under MEBA. Since $t^*$ is the only mode-switch in this busy interval, we can conclude that $t^*$ is the earliest time instant when these conditions are satisfied.
\end{IEEEproof}

Finally, the following lemma shows that for a given $\beta^*$, MEBA is optimal in terms of its ability to postpone the mode-switch when compared to any other offline or runtime strategy that uses a fixed budget for each HC task in the LC mode.

\begin{lemma}
For the dynamic MC task system $\tau$, consider any budget allocation $B_i^L (\forall \tau_i\in\tau_H)$ satisfying the condition $\sum_{\tau_i \in \tau_H} B_i^L/T_i \leq \beta^* \times U_H$, such that each job of each HC task $\tau_i$ is given a budget of exactly $B_i^L$ in the LC mode. Then, for any job sequence (release time and execution demand) of $\tau$, the mode-switch instant based on this budget allocation is no later than the mode-switch instant under MEBA. 
\label{lem:opt}
\end{lemma}
\begin{IEEEproof}
We prove this lemma by contradiction. Suppose under MEBA the mode-switch happens at $t^*$ for some job sequence. From Lemma~\ref{lem:ms} we know that $\sum_{\tau_i\in\tau_H} e_{i,t^*}^m/T_i=\beta^*\times U_H$, where $e_{i,t^*}^m$ denotes the maximum amount of time for which any job of $\tau_i$ has executed before $t^*$. Also, there is at least one HC task $\tau_k$ having an incomplete job at $t^*$ with execution demand $e_k > e_{k,t^*}^m$.

Suppose there exists a budget allocation satisfying the condition $\sum_{\tau_i\in\tau_H} B_i^L/T_i \leq \beta^* \times U_H$ that can further postpone the mode-switch. Then it must be the case that $\forall \tau_i\in \tau_H:~B_i^L\geq e_{i,t^*}^m$ and $B_k^L> e_{k,t^*}^m$. However this is impossible because the assumption $\sum_{\tau_i\in\tau_H} B_i^L/T_i \leq \beta^* \times U_H$ is violated in that case.
\end{IEEEproof}

\subsection{EDF-UVD Scheduling Strategy}
\label{sec:schedule}

We propose a new scheduling strategy for the dynamic model called EDF-UVD (Earliest Deadline First with Universal Virtual Deadlines). EDF-UVD is based on the well known algorithm EDF-VD~\cite{BBA12} in which HC jobs are assigned virtual deadlines shorter than their original deadlines in the LC mode. These virtual deadlines ensure that when a mode-switch occurs, there is sufficient time for HC jobs to complete any additional execution before their original deadlines. The only difference between EDF-VD and EDF-UVD is that under EDF-UVD, even LC jobs would be assigned virtual deadlines shorter than their original deadlines in the LC mode. A job of LC task $\tau_i$ would be scheduled based on its virtual deadline as long as it has not executed for more than $\alpha_i \times C_i$ time units (i.e., execution proportional to its HC service level). Thereafter, it will be scheduled using its original deadline. Thus, EDF-UVD can be formally defined as follows. 
\begin{enumerate}
 	\item Let $D_i(L)=x\times T_i$, where $x\in[0,1]$, denote the virtual deadline for each task $\tau_i \in \tau$. 
	\item When the system is in the LC mode:
	\begin{itemize}
		  \item A job of HC task $\tau_i$ will be scheduled based on its virtual deadline $D_i(L)$.
		  \item A job of LC task $\tau_i$ that has executed for less than $\alpha_i \times C_i$ time units will be scheduled based on its virtual deadline $D_i(L)$.
		  \item A job of LC task $\tau_i$ that has executed for $\alpha_i \times C_i$ time units or more will be scheduled based on its original deadline $D_i$.
	\end{itemize}
	\item When the system is in the HC mode, both LC and HC jobs will be scheduled using their original deadlines.
	\item In each mode, all jobs will be scheduled using the Earliest Deadline First policy.
\end{enumerate}

EDF-UVD can reduce ``unnecessary'' budget allocations for LC tasks in comparison to EDF-VD. This happens when some job of a LC task $\tau_i$ executes beyond $\alpha_i \times C_i$ time units in the LC mode, and the system switches mode before the deadline of this job. In this case, any execution of the job beyond $\alpha_i \times C_i$ is unnecessary, because it is not required to be satisfied based on the definition of MC-schedulable. However, it is impossible to know prior to the mode-switch which execution is unnecessary. To address this issue, we use virtual deadlines for LC jobs in the LC mode. The main intuition behind this strategy is that the first $\alpha_i \times C_i$ execution units would be scheduled with a higher priority (based on virtual deadline), and the remaining execution units would be scheduled with a normal priority (based on original deadline). Thus, all jobs of LC tasks would prioritize the first $\alpha_i \times C_i$ execution units, thereby completing those execution units earlier than in the case of EDF-VD. Below we use a simple example to illustrate how EDF-UVD reduces unnecessary budget allocations.

\begin{example}
\label{example:1}
As shown in Figure~(a), there are two LC jobs $J_1^*$ and $J_2^*$ that are released before $t^*$ but have deadline after $t^*$, where $t^*$ denotes the mode-switch instant. If $J_1^*$ receives $C_1$ units of budget before $t^*$, then deadline miss would happen because $J_2^*$ receives less than $\alpha_2\times C_2$ units of budget before its deadline $D_2$. However as shown in Figure~(b), after $J_1^*$ executes for $\alpha_1\times C_1$ units of time, it is scheduled using its true deadline $T_1$, and at that time $J_2^*$ would have higher priority than $J_1^*$. As a result, the  $(1-\alpha_1)\times C_1$ units of budget previously consumed by $J_1^*$ is now consumed by $J_2^*$, and the deadline miss at $D_2$ is avoided.
\vspace{-2mm}
\begin{figure}[th!]
\centering
\subfigure[$J_1^*$ receives unnecessary budget]{
\label{fig:e11}
        \begin{minipage}[t]{0.3\textwidth}
        \includegraphics[width=\textwidth]{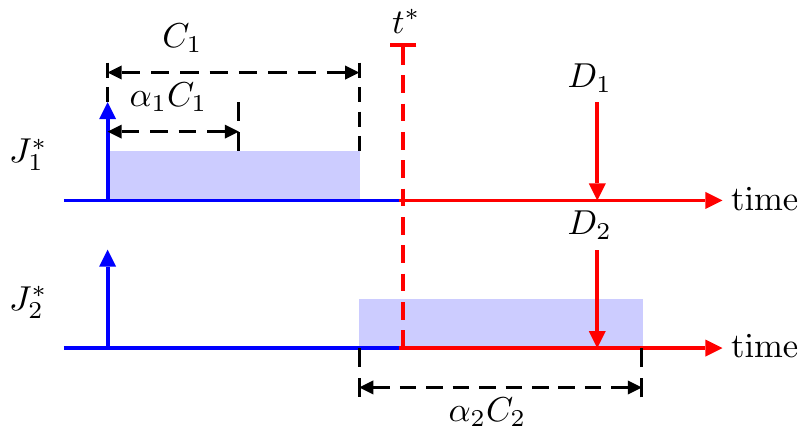}
        \end{minipage}
}
\subfigure[$J_1^*$ does not receive unnecessary budget]{
\label{fig:e12}
        \begin{minipage}[t]{0.3\textwidth}
        \includegraphics[width=\textwidth]{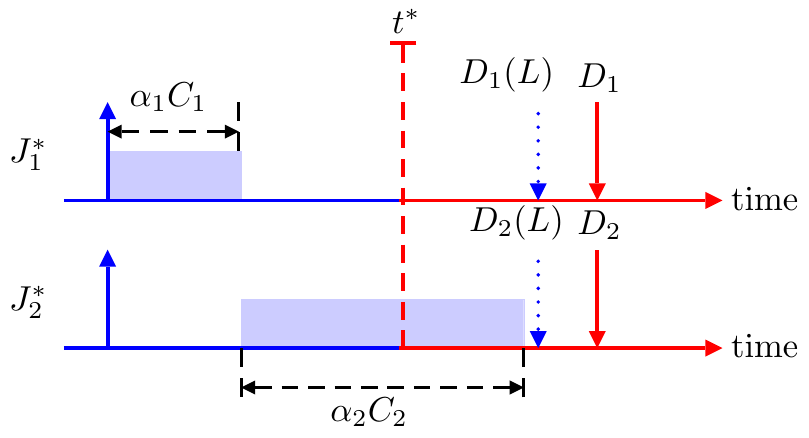}
        \end{minipage}
}
\caption{Example in which EDF-UVD performs better than EDF-VD}
\label{fig:e1}
\end{figure}
\end{example}

\textbf{Runtime complexity for MEBA and EDF-UVD}: For the static model consisting of $n$ tasks, EDF-VD can be implemented efficiently with a runtime complexity of $O(\log n)$ per event, where an event is either the arrival of a job, preemption of a job, completion of a job, or the mode-switch instant. For our proposed dynamic scheduling model (MEBA with EDF-UVD), the additional operations are summarized as follows: 1) When a job of LC task $\tau_i$ executes for $\alpha_i\times C_i$ time units, the scheduler will be invoked to change the job's deadline; 2) $b_i$ is updated whenever a job of HC task $\tau_i$ is allocated to the processor; 3) $e_i^m$ is updated whenever a job of HC task $\tau_i$ completes or is preempted. Therefore, the dynamic scheduling model can also be implemented with $O(\log n)$ runtime complexity per event, but each LC job would generate one additional event when compared to the static model. 

\section{Schedulability Test}
\label{sec:test}

In this section, we derive a sufficient schedulability test for the dynamic MC model. The derived test depends on the values of system service levels $\alpha^*$ and $\beta^*$ (see {$\alpha^*$ and $\beta^*$  in Definition~\ref{def:def2}})\footnote{Techniques for determining $\alpha^*$ and $\beta^*$ are presented in Section~\ref{sec:opt}.}. 

\begin{theorem}
\label{th:1}
	Given $\alpha^*$ and $\beta^*$, a dynamic MC task system $\tau$ is MC-schedulable by MEBA and EDF-UVD on an uniprocessor platform if the following condition holds:
	\begin{equation}
	\label{eqn:test}
		\left(1-\alpha^*\right)\left(1-\beta^*\right)\geq \frac{U_H+U_L-1}{U_L\times U_H},
	\end{equation}
	and the virtual deadline factor $x$ falls in the following range:
\begin{align*}
\left [\frac{\beta^*\times U_H+\alpha^*\times U_L}{1-U_L \times \left(1-\alpha^*\right)},~ \frac{1-U_H-\alpha^*\times U_L}{(1-\alpha^*)\times U_L} \right]
\end{align*}	
\end{theorem}
\begin{IEEEproof} 
We prove this theorem by mapping the dynamic MC task system $\tau$ scheduled by MEBA and EDF-UVD to a static MC task system $\tau'$ scheduled by EDF-VD. Consider a job sequence $\mathcal{J}$ (set of release times and execution demand) of $\tau$ that results in a mode-switch at some time instant $t^*$. Let $e_{i,t^*}^m$ denote the maximum amount of time for which any job of HC task $\tau_i \in \tau$ has executed up to instant $t^*$ in the busy interval containing $t^*$. 

Consider the mapping of tasks from the dynamic model to the static model shown in Table~\ref{tab:notation}. For a LC task $\tau_i=(T_i,C_i,LC)$ with virtual deadline $D_i(L)=T_i\times x$, we map it to a HC task $\tau'_{i,1}$ and a LC $\tau'_{i,2}$. For a HC task $\tau_i=(T_i,C_i,HC)$ with virtual deadline $D_i(L)=T_i\times x$, we map it to a HC task $\tau'_i$.

	\begin{table}[h]
\center
\begin{tabular}{|c|c|c|c|c|c|}
 \hline
Task &  $T_i$& $C_i^L$ & $C_i$ &$L_i$ & $D_i(L)$\\
 \hline
 $\tau'_{i,1}$& $T_i$& $\alpha_i \times C_i$ & $\alpha_i \times C_i$&HC& $T_i\times x$\\
 \hline
 $\tau'_{i,2}$&$T_i$ & - &  $(1-\alpha_i)\times C_i$ & LC & -\\
 \hline
 $\tau'_i$&$T_i$ & $e_{i,t^*}^m$ &  $C_i$ & HC & $T_i\times x$\\
 \hline
\end{tabular}
\caption{Tasks in static MC system $\tau'$}
\label{tab:notation}
\end{table}

Let $\tau' = \{ \tau'_{i,1}, \tau'_{i,2} | \forall \tau_i \in \tau_L \} \bigcup \{ \tau'_i | \forall \tau_i \in \tau_H \}$. Lemma~\ref{cor:mapping} in Appendix~\ref{app:proof} shows that for the job sequence $\mathcal{J}$ generated by the dynamic system $\tau$, it is feasible to generate an \emph{identical} job sequence $\mathcal{J}'$ by the static system $\tau'$. Since the runtime scheduling policy of EDF-UVD and EDF-VD are also identical, if $\tau'$ is MC-Schedulable under EDF-VD, then job sequence $\mathcal{J}$ of $\tau$ is also guaranteed to be MC-Schedulable under MEBA and EDF-UVD. Thus, we can use the schedulability test for EDF-VD~\cite{BBA12} to derive a test for MEBA and EDF-UVD.

{From Definition~\ref{def:def2} we have  $\sum_{\tau_i\in \tau_L} \alpha_i\times \frac{C_i}{T_i}=U_L\times \alpha^*$}, and hence the total utilization of LC tasks in $\tau'$ (tasks of type {$\tau'_{i,2}$} in Table~\ref{tab:notation}) is
$\sum_{\tau_{i}\in \tau_L}  \frac{(1-\alpha_i)\times C_i}{T_i}=U_L\times(1-\alpha^*)$.

The total utilization of HC tasks in $\tau'$ (tasks of type {$\tau'_{i,1}$} and $\tau'_i$ in Table~\ref{tab:notation}) in the HC mode is \[
\sum_{\tau_i\in\tau_H}\frac{C_i}{T_i} + \sum_{\tau_i\in \tau_L} \frac{ \alpha_i\times   C_i}{T_i}=U_H+U_L\times \alpha^*.
\]
{From Lemma~\ref{lem:ms} we have $\sum_{\tau_i\in \tau_H} \frac{e_{i,t^*}^m}{T_i}\leq U_H\times \beta^*$.} Thus the total utilization of HC tasks in $\tau'$ in the LC mode  is \[
\sum_{\tau_i\in \tau_H} \frac{e_{i,t^*}^m}{T_i}+\sum_{\tau_i\in \tau_L} \alpha_i\times \frac{C_i}{T_i} =U_H\times \beta^*+U_L\times \alpha^* .\]

Then, using Theorems 1 and 2 from~\cite{BBA12}, we get that $\tau'$ is MC-Schedulable by EDF-VD on an uniprocessor platform with virtual deadline factor $x$, if
\begin{align*}
& x \leq  \frac{1-U_H-\alpha^*\times U_L}{(1-\alpha^*)\times U_L} \wedge x\geq \frac{\beta^*\times U_H+\alpha^*\times U_L}{1-U_L \times (1-\alpha^*)} \\
& \mbox{and } \left(1-\alpha^*\right)\left(1-\beta^*\right)\geq \frac{U_H+U_L-1}{U_L\times U_H}.
\end{align*}

This proves the theorem because above equations are independent of any parameters specific to $\mathcal{J}$.
\end{IEEEproof}

\section{Determination of System Service Levels}
\label{sec:opt}

From Theorem~\ref{th:1} in Section~\ref{sec:test} we know that a dynamic system $\tau$ is MC-Schedulable under MEBA and EDF-UVD if $(1-\alpha^*)(1-\beta^*)\geq M\mbox{~where~} M=\frac{U_H+U_L-1}{U_L\times U_H}$.

We can see that there is a trade-off between the LC system service level $\beta^*$ and the HC system service level $\alpha^*$. A higher value for $\beta^*$ implies a higher total budget allocation for HC tasks in the LC mode, and consequently the mode-switch can get delayed. It is then possible that LC tasks would continue to receive a higher service level ($=1$) in the LC mode for a longer duration of time. On the other hand, a higher value for $\alpha^*$ implies higher service level for LC tasks in the HC mode, but then the resulting smaller value for $\beta^*$ may lead to a mode-switch at an earlier time instant. Thus, $\beta^*$ controls the amount of time for which LC tasks receive full service, whereas $\alpha^*$ controls the minimum guaranteed service for LC tasks at all times.

If we want to minimize the likelihood that the system switches to HC mode, then we can set $\alpha^*=0$ and choose the maximum possible value for $\beta^* (=1-M)$. Alternatively, if we wish to support LC tasks as much as possible, while ensuring that the likelihood of mode-switch is relatively small, then we can set $\beta^*=\frac{\sum_{\tau_i\in \tau_H}C_i^L/T_i}{U_H}$, where $C_i^L$ denotes the maximum observed execution time for HC task $\tau_i$ in large-scale simulations.


Note that the value of $\beta^*$ is upper bounded by $1-M$. Therefore, if $\tau$ has a large $U_H+U_L$ value (small $1-M$)   { and $1-M\ll \frac{\sum_{\tau_i\in \tau_H}C_i^L/T_i}{U_H}$, then the overall demand of HC tasks will frequently exceed $(1-M)U_H$}. As a result, the system is  likely to switch to HC mode frequently. To support LC tasks as much as possible in this case, it might be better to set $\beta^*=0$ and $\alpha^*=1-M$. 


\textbf{HC budget allocation for LC tasks.} Given $\alpha^*$, we have the flexibility to assign different HC budgets for each LC task $\tau_i$, $B_i^H (=\alpha_i \times C_i)$, as long as $
{\sum_{\tau_i\in \tau_L}\frac{B_i^H}{T_i}}=\sum_{\tau_i\in \tau_L}\frac{\alpha_i\times C_i}{T_i} \leq \alpha^*\times U_L
$.
A simple strategy in which we do not differentiate between LC tasks is to distribute the budget equally, i.e.,  $\forall \tau_i,~\tau_j\in\tau_L:~B_i^H=\frac{T_i}{T_j}B_j^H$. Another possible solution is that the application designer can provide a range of HC budget values for the LC tasks, and the system designer can choose among them based on $\alpha^*$ and the requirements of other LC tasks.

\subsection{Total System Utilization}
\label{sec:wsu}
\begin{figure}[tbh]
\centering
        \includegraphics[width=0.5\textwidth]{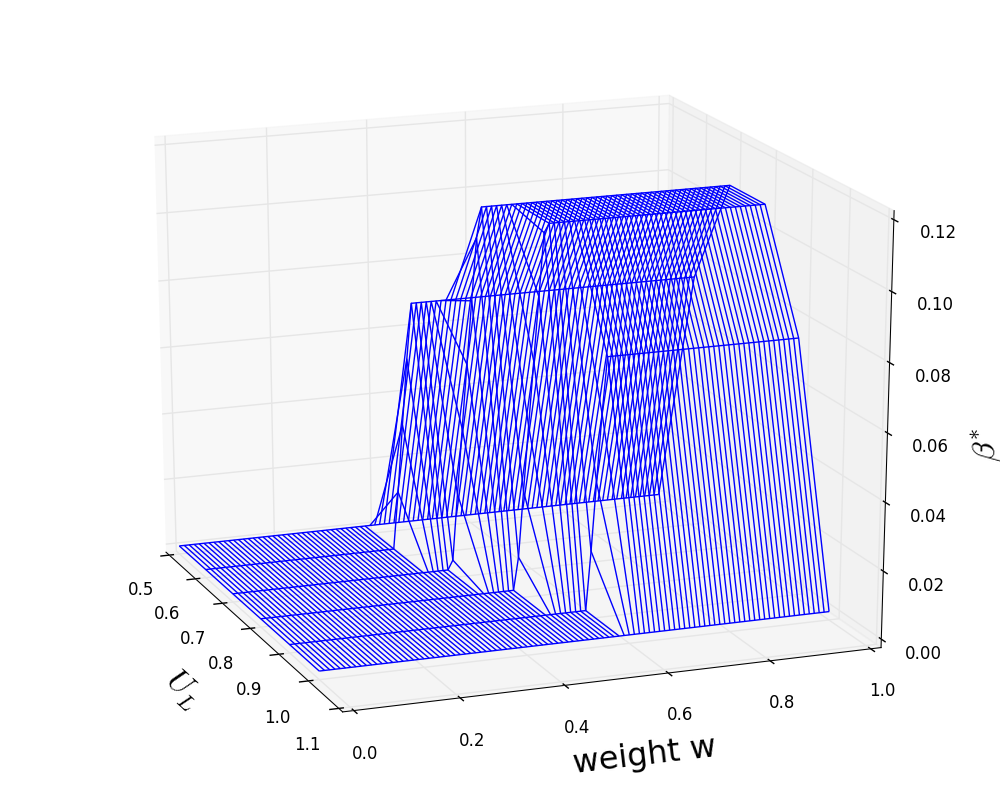}
\vspace{-5mm}
\caption{Relation between $\beta^*$ and weight $w$ when $SU(w)$ is maximized }
\vspace{-5mm}
\label{fig:15}
\end{figure}

When the system is in the LC mode, each LC task $\tau_i$ receives a budget of $C_i$ and all HC tasks combined receive a total budget proportional to $\beta^* \times U_H$. Thus the system utilization in the LC mode can be defined as follows.
\vspace{-2mm}
    \[SU_L=\beta^*\times U_H+\sum_{\tau_i\in \tau_L}\frac{C_i}{T_i} = \beta^*\times U_H+U_L\]
\vspace{-2mm}

Similarly, when the system is in the HC mode, LC tasks receive a budget of $B_i^H=\alpha_i\times C_i$ and HC tasks receive a budget of $C_i$ per job. Thus the system utilization in the HC mode can be defined as follows.
\vspace{-1mm}
\[SU_H=\sum_{\tau_i\in \tau_L}\frac{B_i^H}{T_i}+U_H=\alpha^*\times UL+U_H\]
\vspace{-2mm}

Consider the following definition of \emph{total system utilization}, where $w \in [0,1]$
\vspace{-2mm}
  \begin{equation}
  \begin{split}
      SU(w)&=w\times SU_L+(1-w)\times SU_H\\
        &\leq U_H(1-w+w\beta^*)+U_L(w+(1-w)(1- \frac{M}{1-\beta^*}))\\
        &=U_H(1-w)+U_L+U_Hw\beta^*-\frac{U_L(1-w)M}{1-\beta^*}    
  \end{split}
\end{equation}  
\vspace{-5mm}

$SU(w)$ gives a value for the total system utilization, assuming a weight of $w$ for the LC mode utilization. Since  {$SU(w)$ is a convex function of $\beta^*$ and $\beta^*$ is in the range $[0,1-M]$, by taking the first derivate and equating it to $0$}, $SU(w)$ is maximized when
\vspace{-2mm}
\begin{equation}
\label{eqn:mweight}
    \beta^*=\max\left\{0,\min\left\{1-M, 1-\sqrt{\frac{M(1-w)U_L}{wU_H}}\right\}\right\}
\end{equation}

We performed some experiments to understand the impact of weight $w$ on the total system utilization. In Figures~\ref{fig:15} we show how the value of $\beta^*$ varies when $SU(w)$ is maximized. We fix $U_{sum}=U_H+U_L = 1.5$ and assign different values to $U_L$ and $w$. Z-axis plots the value of $\beta^*$ for which $SU(w)$ is maximized, x-axis denotes $U_L$, and y-axis denotes weight $w$. $U_L\in\{U_{sum}-1,U_{sum}-0.9,U_{sum}-0.8,\ldots,1\}$ and $w\in\{0.02,0.04,0.06,\ldots,1\}$ in the figure. 

As we can observe, in many cases, the value of $\beta^*$ for which $SU(w)$ is maximized is not sensitive to the value of $w$. The reason for this observation is that when the system has a large $U_{sum}$, the value of $1-M$ is relatively small. For such systems, $SU(w)$ is maximized when either $\beta^*=1-M$ or $\beta^*=0$ for almost all the values of $w$.



\section{Evaluation}
\label{sec:eva}

An important advantage of the dynamic model is that  the designer is not required to specify LC budgets for individual HC tasks. There are two other MC models that do not require the designer to provide such budgets. In the worst-case reservations model commonly used in the safety-critical industry today, tasks are always guaranteed to receive $C_i$ units of budget per job, irrespective of their criticality level. The dynamic model generalizes this model, and they are equivalent if we set $\beta^*=1$. Although this model ensures isolation between tasks at different criticality levels, it does not allow for efficient sharing of the processor. Another one is the elastic model~\cite{Su13}, in which all HC tasks receive $C_i$ units of budget per job, and the service level of LC tasks is fixed at $(1-U_H)/U_L$. The dynamic model also generalizes this model, and they are equivalent if we set $\beta^*=0$ and $\alpha^*=(1-U_H)/U_L$. In the remainder of this section, we evaluate the dynamic model from three different aspects.

 
\subsection{Minimum guaranteed service for LC tasks} 
\label{sec:gsl}

In this section we experimentally evaluate the performance of the dynamic model in terms of its ability to provide minimum guaranteed service to LC tasks. We compare its performance to the service adaption strategy~\cite{PCH14}, which  also uses a EDF-based scheduling policy and provides a minimum guaranteed service to LC tasks. This strategy uses EDF-VD and decreases the dispatch frequency of LC tasks when the system switches mode. It finds  the minimal possible factor $y$ (single value)  to extend the periods of all LC tasks i.e., $T_i\leftarrow y\times T_i$, where $y\geq 1$. Therefore, the minimum guaranteed service for LC tasks that the service adaption strategy can support is equal to $\frac{1}{y}$. We compare this value with $\alpha^*$ in the dynamic model to evaluate their relative performance.  

For this comparison, we have to use task systems characterized by the static model (i.e., given $C_i^L$ value for each HC task $\tau_i$), because the service adaptation strategy is designed for such systems. To ensure a fair comparison, we assume that $\beta^*$ in the dynamic model is set to a value such that $\beta^* \times U_H = \sum_{\tau_i\in\tau_H} C_i^L/T_i$. We then compute the maximum possible $\alpha^*$ that still guarantees schedulability.

We use the same task set generation procedure as in \cite{EkYi12}, and it can be summarized as follows. Each task is generated based on the following parameters.
\begin{itemize}
\label{set:1}
    \item Task $\tau_i$ is a HC task with probability $PH=0.5$.
    \item $C_i^L$ is drawn using an uniform distribution over $[1,10]$.
    \item $C_i$ is drawn using an uniform distribution over $\left[C_i^L,~RC\times C_i^L\right]$, where $RC\in \{3,4,5\}$.
    \item $T_i$ is drawn using an uniform distribution over $\left[C_i,200\right]$.
\end{itemize}

Task set $\tau$ is empty initially. Randomly generated tasks based on the above procedure are added to the task set repeatedly. Let $U_A=\frac{U_L+U_H+U_H\beta^*}{2}$ denote the average utilization of task set $\tau$ in LC and HC mode at any point in the generation process (computed based on tasks that have already been added to $\tau$). We classify $\tau$ based on the range in which $U_A$ lies: $[0.54,0.55]$, $[0.59,0.60]$, $[0.64,0.65]$, $[0.69,0.7]$, $[0.74,0.75]$. We do not consider the case when $U_A>0.75$, because very few task sets with $U_A>0.75$ are schedulable under the service adaption strategy. A new task is added to $\tau$ until $U_A$ falls in the range we choose. However if $U_{A}$ becomes greater than the upper bound of the range, we discard the entire task set and repeat the process. 


\begin{table}[h]
\caption{{Average minimum guaranteed service for LC tasks}}
\label{fig:serc3}
\footnotesize
\center
\begin{tabular}{|l|l|l|l|l|l|l|}
  \hline
Average Utilization &0.55&0.6&0.65&0.7&0.75\\
 \hline
Dynamic Model RC=3&  $0.985$& $0.931$ & $0.832$&$0.566$&$0.235$\\
 \hline
Service Adaption RC=3 & $0.976$ &$0.882$& $0.636$&$0.268$&$0.177$ \\
 \hline
Dynamic Model RC=4   &$0.988$&$0.950$&$0.831$&$0.643$&$0.321$\\
 \hline
Service Adaption RC=4  &$0.984$ &$0.903$& $0.639$ &$0.326$&$0.129$\\
\hline
Dynamic Model RC=5  &$0.978$&$0.912$&$0.648$&$0.295$&$0.089$\\
 \hline
Service Adaption RC=5  &$0.964$ &$0.805$& $0.339$ &$0.210$&$0.053$\\
\hline
\end{tabular}
\end{table}
{In Table~\ref{fig:serc3},  we show the average minimum guaranteed service for LC tasks that the dynamic model and service adaption strategy can support when $RC=3,~4$ and $5$.} Each value  in the table is based on $1000$ task sets. If a task set is not schedulable by either the dynamic model or the service adaptation strategy, then the service level of LC tasks is assumed to be $0$. This is reasonable because there is no guarantee on deadlines for such task sets. Even though both the service adaption strategy and the dynamic model are based on EDF, the dynamic model always outperforms the service adaption strategy. For some settings, its performance is almost two times better than the performance of the latter.


The reasons resulting in this performance gap can be summarized as follows: 1) the schedulability analysis for service adaption strategy is based on pessimistic approximate demand bound functions whereas for the dynamic model it is based on utilization-based tests, and 2) we set virtual deadlines for LC tasks to reduce the scenario where jobs of LC tasks receive unnecessary budget (advantage of EDF-UVD over EDF-VD). 
Another important advantage of the dynamic model is that it provides the flexibility to set different HC service levels for each LC task ($\alpha^*$ can be split among the different LC tasks based on $\alpha_i$), while the service adaption strategy has to decrease the dispatch frequency of all the LC tasks to the same degree. 


\subsection{Analysis of probability of mode-switch}
\label{sec:apms}
Under the dynamic model the mode-switch is triggered only when the total allocated LC budget for all HC tasks combined exceeds $\beta^* \times U_H$. This is different from the static model in which a mode-switch is triggered even when a single HC task exceeds its allocated budget. Therefore, the dynamic model can reduce the probability that a task system switches mode when compared to the static model. In this section we analytically compare the two models from this aspect. We first use a simple example to illustrate how the dynamic model can reduce the likelihood of mode-switch.


\begin{example}
Suppose we have a task system $\tau=\{\tau_1,\tau_2,\tau_3\}$, where $\tau_1=(10,5,LC),~\tau_2=(10,4,HC),~\tau_3=(10,4,HC)$. Using Equation~\ref{eqn:test}, we set $\beta^*=0.25$ and $\alpha^*=0$. Thus, as long as $(e_{1}^m/10+e_{2}^m/10)/0.8\leq 0.25$, i.e., $e_{1}^m+e_{2}^m\leq 2$, the system would stay in the LC mode (from Lemma~\ref{lem:ms}). Instead, if we have a fixed budget as in the static model, e.g., $B_1^L=B_2^L=1$, the system would switch to HC mode when either $\tau_2$ or $\tau_3$ executes beyond $1$ time unit. Thus, in the dynamic model, even if $\tau_2$ executes for $1.05$ time units, the system could stay in the LC mode as long as no job of $\tau_3$ has executed beyond $0.95$ time units. 
\end{example}

Let $P(s_i)$ denote the probability that no job of $\tau_i$ executes for more than $C_i\times s_i$ time units in a certain  busy interval, {where $s_i$ is a random variable in the range   $(0,1]$}. Then, a task set $\tau$ does not switch mode under the static model only when no HC task executes beyond its fixed LC budget $B_i^L = C_i^L$ determined offline (let $C_i^L=\beta_i\times C_i$ for each $i$). Then, assuming all tasks are independent, the probability that the system does not switch to HC mode is
\begin{equation}
\label{eq:psw}
P_{noswitch}^s=\prod_{\tau_i\in \tau_H} P(s_i=\beta_i)
\end{equation}

On the other hand, {according to Lemma~\ref{lemma:mode-switch}, the system does not switch mode under the dynamic model as long as $\sum_{\tau_i\in \tau_H} \frac{e_{i}^m}{T_i}\leq \beta^*\times U_H$. Suppose $e_{i}^m=s_i\times C_i$. Then, } the probability that the system does not switch to HC mode under the dynamic model is
\begin{equation}
\label{eq:psw1}
P_{noswitch}^d=\sum\limits_{\sum\limits_{\tau_i\in \tau_H} \frac{s_i\times C_i}{T_i}\leq \beta^*\times U_H }^{s_i\in(0,1]}\left(\prod_{\tau_i\in \tau_H} P(s_i)\right)
\end{equation}

We can see that $P_{noswitch}^d\geq P_{noswitch}^s$, because $s_i=\beta_i$ is one possible assignment that satisfies $\sum_{\tau_i\in \tau_H} \frac{s_i\times C_i}{T_i}\leq \beta^*\times U_H$ (using Lemma~\ref{lem:opt}), but there exists many more assignments of $s_i$  that also satisfy the above condition. Further, we can also observe that the probability of mode-switch under the static model increases exponentially with the number of HC tasks, while the dynamic model can mitigate this problem. 


\begin{table}[h]
\caption{{Example distribution for $P(s_i)$}}
\label{fig:distA}
\small
\center
\begin{tabular}{|l|l|l|l|l|l|}
 \hline
$s_i$&  $0.1$& $0.2$ & $0.3$&$0.4$&$0.5$\\
 \hline
$P(s_i)$  & $0.01$ &$0.05$& $0.2$&$0.5$&$0.8$ \\
 \hline
 $s_i$   &$0.6$&$0.7$&$0.8$&$0.9$&$1.0$\\
 \hline
 $P(s_i)$  &$0.9$ &$0.95$& $0.98$ &$0.995$&$1.0$\\
\hline
\end{tabular}
\end{table}

{From Equations~\ref{eq:psw} and \ref{eq:psw1}, we can evaluate the performance of dynamic and static models analytically in terms of their  ability to reduce the mode-switch probability.}  Now  we plot  Equation~\ref{eq:psw} and \ref{eq:psw1} using a specific distribution for $P(s_i)$. We set $\beta^*=1-M=(1-U_L)(1-U_H)/(U_L.U_H)$ and $\alpha^*=0$ in the dynamic model because this choice postpones the mode-switch as much as possible. For the static model, to ensure a fair comparison, we set $\beta_i=1-M\Leftarrow C_i^L=(1-M)\times C_i$ for each HC task $\tau_i$. Thus, both the models have the same LC system service level for HC tasks ($\beta^*$). But the dynamic model uses the runtime strategy MEBA to distribute this service among the HC tasks, while the static model uses $\beta_i \times C_i$ as a fixed LC budget.  
\begin{figure}[th!]
\centering
\includegraphics[width=0.43\textwidth]{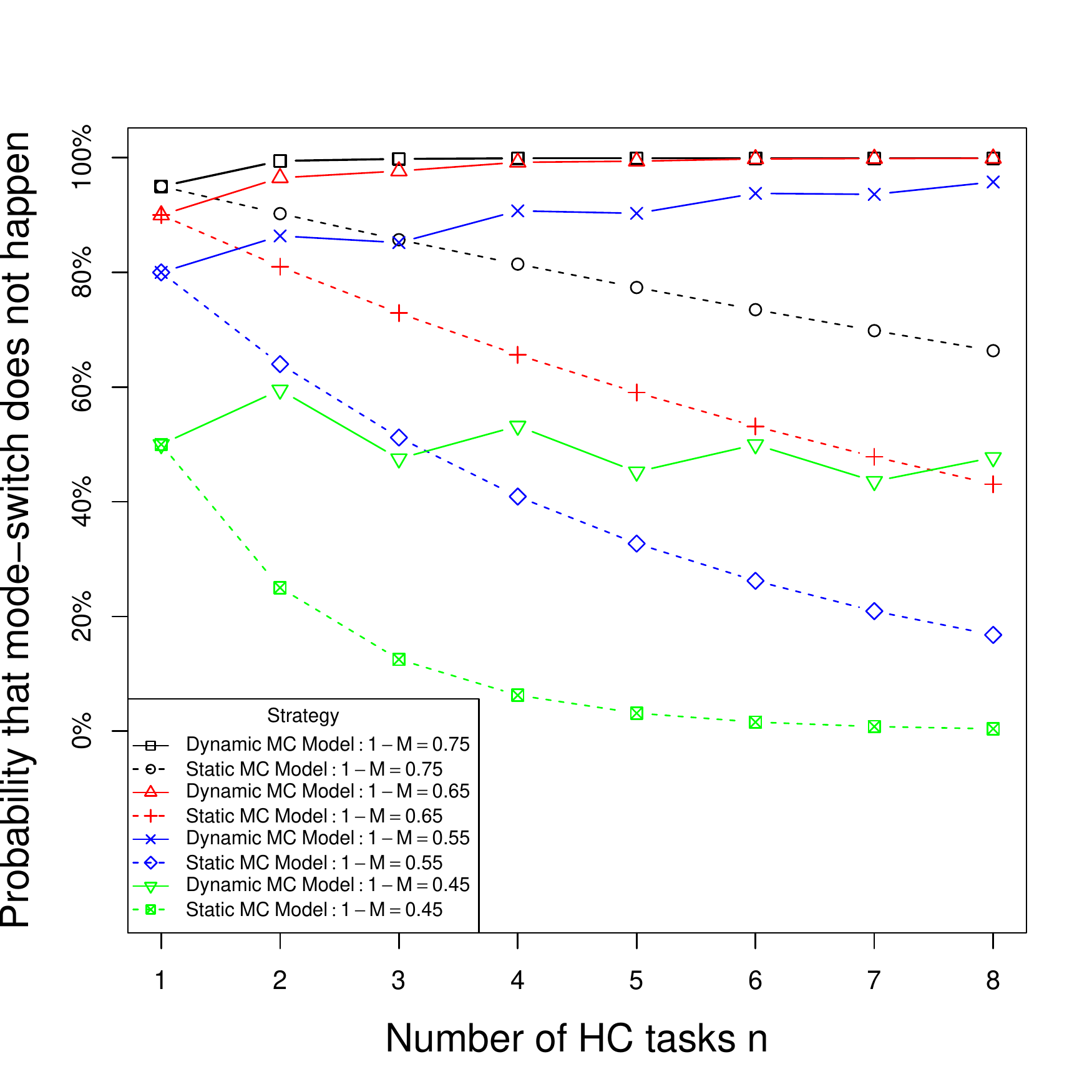}
\caption{Comparison of the probability that the mode-switch does not happen}
\label{fig:resultA}
\end{figure}

Suppose $\tau$ comprises $n$ HC tasks and $P(s_i)$ conforms to the distribution shown in Table~\ref{fig:distA} for each task. Then we can calculate $P_{noswitch}^s$ and $P_{noswitch}^d$ when $n\in\{1,2,3,4,5,6,7,8\}$. Figure~\ref{fig:resultA} plots Equations~\ref{eq:psw} and~\ref{eq:psw1} as a function of the number of HC tasks, where x-axis denotes the number of HC tasks and y-axis denotes the probabilities. The mode-switch probability depends on the value of $\beta^*=1-M$. Hence, we show the results for $1-M\in \{0.45,0.55,0.65,0.75\}$. When $1-M>0.75\vee 1-M<0.45$, the probability would either approach $1$ or $0$ according to the distribution in Figure~\ref{fig:distA}. Hence we do not present the results for $1-M>0.75\vee 1-M<0.45$.

We can see that when $1-M$ has a large value, the mode-switch probability for both dynamic and static models is low, and vice versa. Also, the dynamic model can significantly reduce this probability when compared to the static model under different settings. We can observe that as the number of tasks increase, $P_{noswitch}^s$ decreases  monotonically, while the value of $P_{noswitch}^d$ is stable.

Note that the mode-switch probability for the static model that we computed is in fact the probability of mode-switch in several MC studies~(e.g., \cite{BBD11,BaFo11,GES11,BBA12,EkYi12,Eas13,PCH14,HND14,Weaklyhard,BuBa13,JZP13}). Other works (e.g., \cite{PCH13,GU15,ren2015mixed,fleming2014incorporating}) have different levels of mode-switch depending on the number of HC tasks that are executing beyond their LC budgets; in these studies the first mode-switch has a probability identical to what we computed for the static model. Hence, the results of Figure~\ref{fig:resultA} provide a direct comparison between the dynamic model and all the above studies.

\subsection{Total system utilization}
\label{sec:etsu}
We introduced a technique to maximize the total system utilization in Section~\ref{sec:opt}. Here we compare the maximum total system utilization that dynamic and static models can support for different values of $U_H$ and $U_L$. For the dynamic model  this parameter can be calculated using Equation~\eqref{eqn:mweight}. For the static model, assuming it is scheduled by EDF-VD and no service guarantee for LC tasks after mode-switch, the total system utilization is maximized when $\sum_{\tau_i\in \tau_H}C_i^L/T_i=(1-M)U_H$.
\[SU(w)=w.U_L+w.(1-U_L)(1-U_H)/U_L+(1-w)U_H\] 
Here $w$ denotes the weight for LC mode. Since there is no service guarantee for LC tasks, the utilization is maximized by postponing the mode-switch as much as possible, i.e., similar to setting $\beta^*=1-M$ in the dynamic model.

\begin{figure}[t]
\centering
\includegraphics[width=0.45\textwidth]{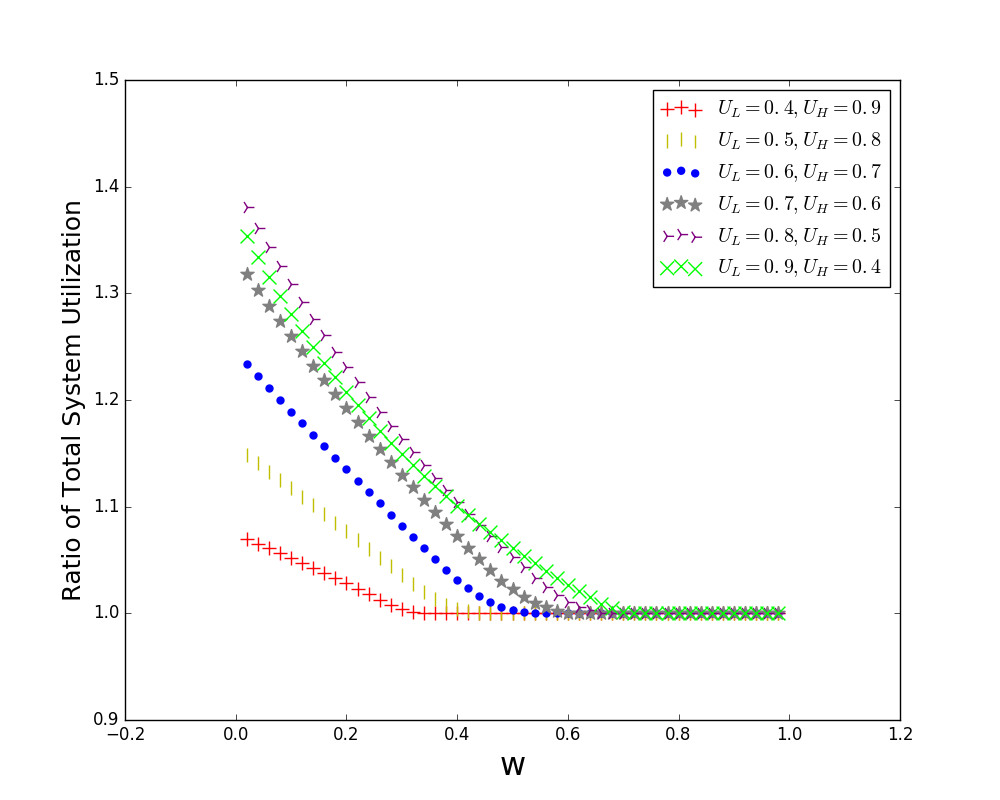}
\caption{Total system utilization when $U_L+U_H=1.3$}
\label{fig:U13}
\vspace{-5mm}
\end{figure}

Figure~\ref{fig:U13} shows the ratios between the total system utilization of dynamic and static models as a function of weight $w\in\{0.02,0.04,\ldots,1.0\}$ when $U_L+U_H=1.3$. As we can observe, when the weight for LC mode is relatively small, there is a clear performance gap between the two models. After the weight $w$ exceeds a certain value, their performance overlaps. This indicates that when the weight $w$ exceeds a certain value, the total system utilization is maximized when $\beta^*=1-M$. That is, if the weight for the LC mode is high enough and therefore we do not care about guaranteed service for LC tasks in the HC mode, then it is best to maximize the LC system service level for HC tasks ($\beta^*$). 

\section*{Acknowledgment}
This work was supported by MoE Tier-2 grant (MOE2013-T2-2-029), Singapore.

\section{Conclusion}
In this paper we proposed a dynamic LC budget allocation mechanism for HC tasks to overcome the limitations of static execution estimates. Unlike the static model where the LC budget of each HC task is required to be provided by the application designer, the dynamic model determines it at runtime based on observed job execution times. The system switches mode in the dynamic model only when the total LC budget allocation for all HC tasks combined is violated.  We also proposed a mechanism that enables LC tasks to receive a minimum guaranteed budget allocation at all times, even in the HC mode. Finally, we presented metrics and explored the trade-off between the total LC budget allocation for HC tasks and the minimum guaranteed service for LC tasks.


In the future, we plan to further explore this trade-off between budget allocation for HC tasks and service guarantee for LC tasks. In particular, we plan to investigate techniques that can use the knowledge of task execution times (e.g., probabilistic worst-case execution time) to maximize the expected system utilization. Another direction of research is improving the runtime strategy for budget allocation to HC tasks. In MEBA we record the maximum execution time among all the jobs in the past (i.e., $e_i^m$) to trigger a mode-switch. However it may be possible to further postpone the mode-switch by using other parameters (e.g., the sum of executions of all the jobs in the past), and we plan to explore such techniques. 

\bibliographystyle{IEEEtran}
\bibliography{References}
\appendices




\section{Proofs}
\label{app:proof}

\begin{lemma}
\label{cor:mapping}
For each job sequence $\mathcal{J}$ generated by the dynamic system $\tau$ in Theorem~\ref{th:1}, it is feasible to generate an \emph{identical} job sequence $\mathcal{J}'$ in the static system $\tau'$ given by Table~\ref{tab:notation}. Here identical means that for each job $J \in \mathcal{J}$, there is a job $J' \in \mathcal{J}'$ such that $J$ and $J'$ have the same release time, execution demand and virtual and original deadlines.
\end{lemma}
\begin{IEEEproof}
First we consider some simplifications to jobs of $\mathcal{J}$. For any $J \in \mathcal{J}$ released after $t^*$ such that $J$ is a job of a LC task $\tau_i$, we assume that its execution demand is bounded by $\alpha_i \times C_i$. This is reasonable because the job cannot execute beyond this bound under EDF-UVD. Also, for any $J \in \mathcal{J}$ released at $t$ such that $J$ is a job of a LC task $\tau_i$, if $J$ executes for more than $\alpha_i \times C_i$ time units (say $e$), then we replace it with two jobs $J_a$ and $J_b$. $J_a$ is released at $t$, has a demand of $\alpha_i \times C_i$ and deadline at $t+xT_i$. $J_b$ is also released at $t$, has a demand of $e_i-\alpha_i \times C_i$ and deadline at $T_i$. Note that under EDF-UVD, $J_b$ will begin execution only after $J_a$ completes, and hence their combined schedule is identical to that of $J$.

Consider the following job sequence in the static model (denoted as $\mathcal{J}'$). For each job $J \in \mathcal{J}$ released at time instant $t$, having execution demand $e$ and deadline at $t'$, a job $J'$ will be released by $\tau'$ with the same parameters, such that:
\begin{enumerate}
\item If $J$ is a job of HC task $\tau_i$, then static HC task $\tau'_i$ will release the job $J'$.
\item If $J$ is a job of LC task $\tau_i$, $t < t^*$ and $J$ has deadline at $t+T_i$ (job type $J_b$), then static HC task $\tau'_{i,2}$ will release the job $J'$.
\item If $J$ is a job of LC task $\tau_i$, and $t \geq t^*$ (job in the HC mode) or $J$ has deadline at $t+xT_i$ (job type $J_a$), then static HC task $\tau'_{i,1}$ will release the job $J'$.
\end{enumerate}

We now show that $\mathcal{J}'$ is valid, that is it is feasible to generate such a sequence. The periods of the mapped tasks in the static model are identical to the periods of the corresponding tasks in the dynamic model. Also, for each job execution demand $e$ in the dynamic model, the corresponding execution demands in the static model are no larger than the task execution time parameters. For case~1 it is easy because $e \leq C_i$. For case~2, $e \leq (1-\alpha_i)C_i$ because it is of type $J_b$. Finally, for case~3, $e \leq \alpha_i \times C_i$ because either it is of type $J_a$ or it is a job released after $t^*$. Similarly, it can be seen from the mapping that the deadlines can also be matched as long as mode-switch is triggered in the static model at $t^*$.

Now we show that the mode-switch is indeed triggered at $t^*$ even in the static model. Observe that the runtime scheduling policy of EDF-UVD and EDF-VD are identical (they both use EDF). Hence, up to time instant $t^*$, since the job sequences in $\mathcal{J}$ and $\mathcal{J'}$ are identical, the schedule is equivalent (whenever a job $J$ is scheduled in the dynamic model, corresponding job $J'$ is scheduled in the static model). This means, at $t^*$, if a job $J \in \mathcal{J}$ has remaining execution time, then job $J' \in \mathcal{J}'$ also has the same amount of remaining execution. From Lemma~\ref{lem:ms} we know that at $t^*$ there is an incomplete job $J$ of a HC task $\tau_i \in \tau$ with $e > e_{i,t^*}^m$, where $e$ denotes the total execution requirement of $J$. Then, we can conclude that $J'$ of HC task $\tau'_i \in \tau'$ also has remaining execution $e' = e > e_{i,t^*}^m$ at $t^*$. Further, by definition we know that for each HC task $\tau_i \in \tau$, no job has executed for more than $e_{i,t^*}^m$ time units before $t^*$. Then, $t^*$ is the first time instant at which any job in $\mathcal{J}'$ is requesting for more execution than its LC budget (since $\mathcal{J}$ and $\mathcal{J}'$ have equivalent schedule until $t^*$). This concludes the proof.     
\end{IEEEproof}

\end{document}